\documentclass[runningheads]{llncs}
\usepackage{graphicx}
\usepackage{hyperref} 
\usepackage{multirow}
\usepackage{graphicx}
\usepackage{amsmath}
\usepackage{amssymb}
\usepackage{booktabs}
\usepackage{multirow}
\usepackage{xcolor}
\usepackage{makecell}
\usepackage{enumitem}
\begin{document}
%
\title{UNeXt: MLP-based Rapid Medical Image Segmentation Network}
%
%
\author{Jeya Maria Jose Valanarasu and Vishal M. Patel } 
%
\institute{Johns Hopkins University}
%
\maketitle              
\begin{abstract}

UNet and its latest extensions like TransUNet have been the leading medical image segmentation methods in recent years. However, these networks cannot be effectively adopted for rapid image segmentation in point-of-care applications as they are parameter-heavy, computationally complex and slow to use.  To this end, we propose UNeXt which is a Convolutional multilayer perceptron (MLP) based network for image segmentation. We design UNeXt in an effective way with an early convolutional stage and a MLP stage in the latent stage. We propose a tokenized MLP block where we efficiently tokenize and project the convolutional features and use MLPs to model the representation. To further boost the performance, we propose shifting the channels of the inputs while feeding in to MLPs so as to focus on learning local dependencies. Using tokenized MLPs in latent space reduces the number of parameters and computational complexity while being able to result in a better representation to help segmentation. The network also consists of skip connections between various levels of encoder and decoder.   We test UNeXt on multiple medical image segmentation datasets and show that we reduce the number of parameters by \textbf{72x}, decrease the computational complexity by \textbf{68x}, and improve the inference speed by \textbf{10x} while also obtaining better segmentation performance over the  state-of-the-art medical image segmentation architectures. Code is available at \href{https://github.com/jeya-maria-jose/UNeXt-pytorch}{https://github.com/jeya-maria-jose/UNeXt-pytorch}

\end{abstract}
\section{Introduction}

Medical imaging solutions have played a pivotal role for diagnosis and treatment in the healthcare sector. One major task in medical imaging applications is segmentation as it is essential for computer-aided diagnosis and image-guided surgery systems. Over the past decade, many works in the  literature have focused on developing efficient and robust segmentation methods. UNet \cite{ronneberger2015u} is a landmark work which showed how efficient an encoder-decoder convolutional network with skip connections can be for medical image segmentation. UNet has became the backbone of almost all the leading methods for medical image segmentation in recent years. Following UNet, a number of key extensions like UNet++ \cite{zhou2018unet++}, UNet3+ \cite{huang2020unet}, 3D UNet \cite{cciccek20163d},  V-Net \cite{milletari2016v}, Y-Net \cite{mehta2018net} and KiUNet \cite{valanarasu2020kiu,valanarasu2021kiu} have been proposed.   Recently, many transformer-based networks have  been proposed for medical image segmentation as they learn a global understanding of images which can be helpful in segmentation. TransUNet \cite{chen2021transunet} modifies the ViT architecture \cite{dosovitskiy2020image}  into an UNet for 2D medical image segmentation. Other transformer-based networks like MedT \cite{jose2021medical}, TransBTS \cite{wang2021transbts}, and UNETR \cite{hatamizadeh2022unetr} have also been proposed for medical image segmentation. Note that almost all the above works have focused on improving the performance of the network but do not focus much on the computational complexity, inference time or the number of parameters, which are essential in many real-world applications. As most of these are used for analysis in laboratory settings, they are tested using machines with high compute power (like GPUs). This helps accelerate the speed of inference and also help accommodate a large number of parameters. 

\begin{figure*}[bp]

	\centering
	\includegraphics[width=1\linewidth]{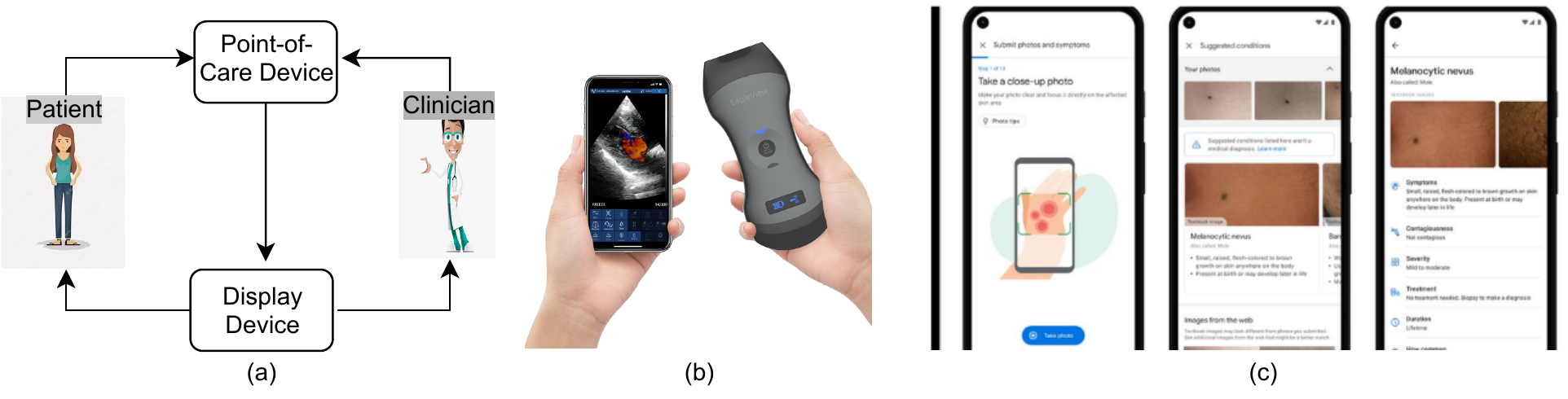}

	\caption{Motivation for UNeXt: As medical imaging solutions become more applicable at point-of-care, it is important to focus on making the deep networks light-weight and fast while also being efficient.  (a) Point-of-Care medical intervention workflow. (b) Recent medical imaging developments: POCUS device \cite{pocus} and (c) Phone-based skin lesion detection and identification application \cite{derm}.}  

	\label{mot}

\end{figure*}

In recent times, there has been a translation of medical imaging solutions from laboratory to bed-side settings. This is termed as point-of-care imaging as the testing and analysis is done by the side of the patient. Point-of-care imaging \cite{vashist2017point} helps clinicians with expanded service options and improved patient care. It helps in reducing the time and procedures involved in patients having to go visit radiology centers. Technology improvements around point-of-care imaging are leading to greater patient satisfaction.  The use of point-of-care devices has been increasing in recent years. For example,  point-of-care ultrasound (POCUS) devices \cite{pocus} have shown to be useful to quickly check  pleural irregularities in lungs, cardiac hemodynamic flow and automatic bladder volume calculation. Phone-camera based images are also being used to detect and diagnose skin conditions \cite{derm}. Magnetic resonance imaging (MRI) machines have also been developed for bed-side operation and fast analysis \cite{mri}.  These recent diagnostic developments have helped in clear and fast acquisition of medical images at point-of-care as seen in Fig. \ref{mot}. Tasks like segmentation, classification and registration are also being integrated along with these appliances to help patients and clinicians accelerate the diagnosis process. The major deep-learning based solutions for these tasks (like UNet and TransUNet) come with an inherent computation overhead and a large number of parameters making them difficult to use in point-of-care applications. In this work, we focus on solving this problem and design an efficient network that has less computational overhead, low number of parameters, a faster inference time while also maintaining a good performance. Designing such a network is essential to suit the shifting trends of medical imaging from laboratory to bed-side. To this end, we propose UNeXt which is designed using convolutional networks and (multilayer perceptron) MLPs.

Recently, MLP-based networks \cite{yu2022s2,touvron2021resmlp,lian2021mlp,tolstikhin2021mlp} have also been found to be competent in computer vision tasks. Especially MLP-Mixer \cite{tolstikhin2021mlp}, an all-MLP based network which gives comparable performance with respect to transformers with less computations. Inspired by these works, we propose UNeXt which is a convolutional and MLP-based network. We still follow a 5-layer deep encoder-decoder architecture of UNet with skip connections but change the design of each block. We have two stages in UNeXt- a convolutional stage followed by an MLP stage. We use convolutional blocks with less number of filters in the initial and final blocks of the network. In the bottleneck, we use a novel Tokenized MLP (Tok-MLP) block which is effective at maintaining less computation while also being able to model a good representation. Tokenized MLP projects the convolutional features into an abstract token and then uses MLPs to learn meaningful information for segmentation. We also introduce shifting operation in the MLPs to extract local information corresponding to different axial shifts. As the tokenized features are of the less dimensions and MLPs are less complicated than convolution or self-attention and transformers; we are able to reduce the number of parameters and computational complexity significantly while also maintaining a good performance. We evaluate UNeXt on ISIC skin lesion dataset \cite{codella2018skin} and Breast UltraSound Images (BUSI) dataset \cite{al2020dataset} and show that it obtains better performance than recent generic segmentation architectures. More importantly, we reduce the number of parameters by \textbf{72x}, decrease the computational complexity by \textbf{68x}  and increase the inference speed by \textbf{10x} when compared to TransUNet making it suitable for point-of-care medical imaging applications. 

In summary, this paper makes the following  contributions: 1) We propose UNeXt, the first convolutional MLP-based network for image segmentation. 2) We propose a novel tokenized MLP block with axial shifts to efficiently learn a good representation at the latent space. 3) We successfully improve the performance on medical image segmentation tasks while having less parameters, high inference speed, and low computational complexity. 

\section{UNeXt}

\noindent \textbf{Network Design:} UNeXt is an encoder-decoder architecture with two stages: 1) Convolutional stage, and a 2) Tokenized MLP stage. The input image is passed through the encoder where the first 3 blocks are convolutional and the next 2 are Tokenized MLP blocks. The decoder has 2 Tokenized MLP blocks followed by 3 convolutional blocks. Each encoder block reduces the feature resolution by 2 and each decoder block increases the feature resolution by 2. Skip connections are also included between the encoder and decoder. The number of channels across each block is a hyperparameter denoted as $C1$ to $C5$. For the experiments using UNeXt architecture, we follow $C1=32$, $C2=64$, $C3=128$, $C4=160$, and $C5=256$ unless stated otherwise. Note that these numbers are actually less than the number of filters of UNet and its variants contributing to the first change to reduce parameters and computation.

\begin{figure*}[]
	\centering
	\includegraphics[width=1\linewidth]{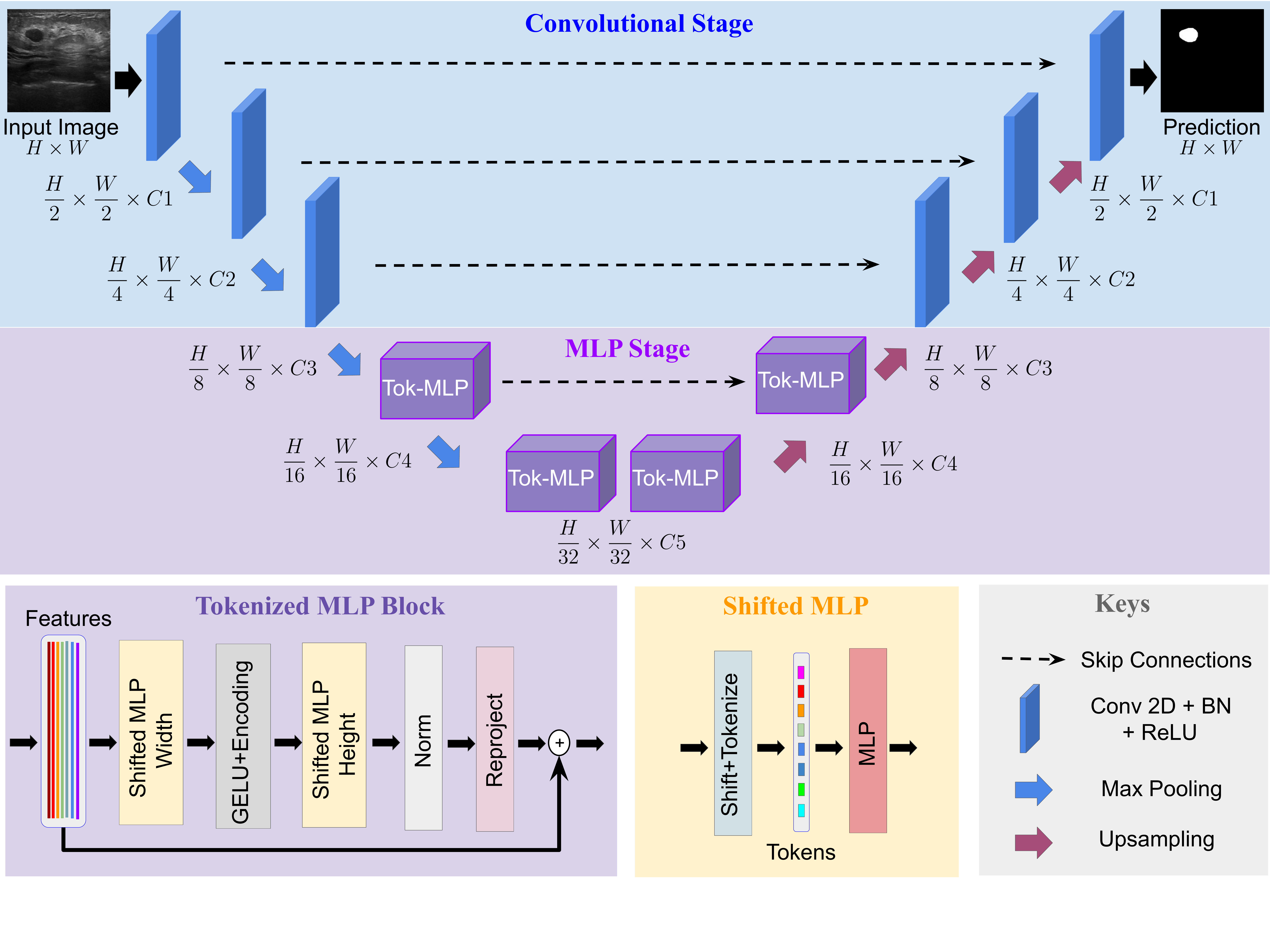}
\vskip -2 em
	\caption{Overview of the proposed UNeXt architecture.}    
	\label{arch}

\end{figure*}
\noindent \textbf{Convolutional Stage:} Each convolutional block is equipped with a convolution layer, a batch normalization layer and ReLU activation. We use a kernel size of $3 \times 3$, stride of $1$ and padding of $1$. The conv blocks in the encoder use a max-pooling layer with pool window $2 \times 2$ while the ones in the decoder consist of a bilinear interpolation layer to upsample the feature maps. We use bilinear interpolation instead of transpose convolution as transpose convolution is basically learnable upsampling and contributes to more learnable parameters.

\noindent \textbf{Shifted MLP:} In shifted MLP, we first shift the axis of the channels of conv features before tokenizing. This helps the MLP to focus on only certain locations of the conv features thus inducing locality to the block. The intuition here is similar to Swin transformer \cite{cao2021swin} where window-based attention is introduced to add more locality to an otherwise completely global model. As the Tokenized MLP block has 2 MLPs, we shift the features across width in one and across height in another like in axial-attention \cite{wang2020axial}. We split the features to $h$ different partitions and shift them by $j$ locations according to the specified axis. This helps us create random windows introducing locality along an axis.

\begin{figure*}[]
	\centering

	\includegraphics[width=0.8\linewidth]{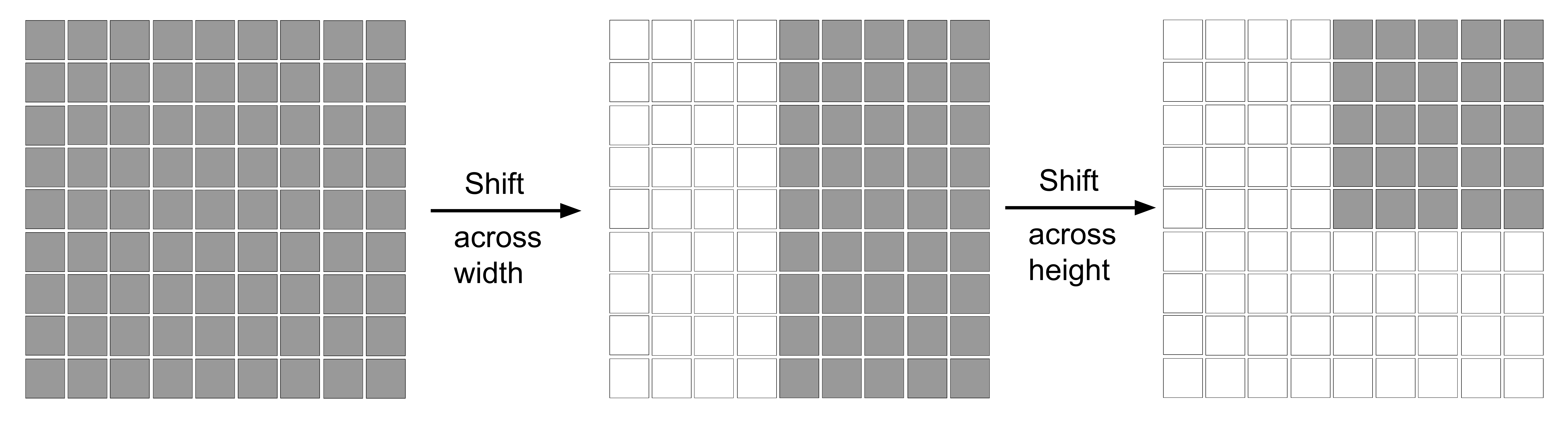}

	\caption{Shifting operation. The features are shifted sequentially across height and width before tokenizing to induce window locality in the network.   }  
	\label{shift}

\end{figure*}
\noindent \textbf{Tokenized MLP Stage:} In the tokenized MLP block, we first shift the features and project them into tokens. To tokenize, we first use a kernel size of 3 and change the number of channels to $E$, where $E$ is the embedding dimension (number of tokens) which is a hyperparameter. We then pass these tokens to a shifted MLP (across width)  where hidden dimensions of the MLP is a hyperparameter $H$. Next, the features are passed through a depth wise convolutional layer (DWConv). We use DWConv in this block for two reasons: 1) It helps to encode a positional information of the MLP features. It is shown in \cite{xie2021segformer} that Conv layer in an MLP block is enough to encode the positional information and it actually performs better than the standard positional encoding techniques. Positional encoding techniques like the ones in ViT need to be interpolated when the test and training  resolutions are not the same often leading to reduced performance. 2) DWConv uses less number of parameters and hence increases efficiency. We then use a GELU \cite{hendrycks2016gaussian} activation layer. We use GELU instead of RELU as it is a more smoother alternative and is found to perform better.  In addition, most recent architectures like ViT \cite{dosovitskiy2020image} and BERT \cite{devlin2018bert} have successfully used GELU to obtain improved results. We then pass the features through another shifted MLP (across height) that converts the dimensions from $H$ to $O$. We use a residual connection here and add the original tokens as residuals. We then apply a layer normalization (LN) and pass the output features to the next block. LN is preferred over BN as it makes more sense to normalize along the tokens instead of normalizing across the batch in the Tokenized MLP block.

The computation in the Tokenized MLP block can be summarized as:
\begin{align}
 X_{shift}  &= Shift_{W}(X); T_{W}  = Tokenize(X_{shift}),\\
 Y &= f(DWConv((MLP(T_{W})))),\\
  Y_{shift}  &= Shift_{H}(Y); T_{H}  = Tokenize(Y_{shift}),\\
  Y &= f(LN(T + MLP(GELU(T_{H})))),
    \end{align}
where $T$ denotes the tokens, $H$ denotes height, $W$ denotes width, $DWConv$ denotes depth-wise convolution and $LN$ denotes layer normalization. Note that all of these computations are performed across the embedding dimension $H$ which is significantly less than the dimensionality of the feature maps $\frac{H}{N} \times \frac{H}{N}$ where $N$ is a factor of 2 depending on the block. In our experiments, we set $H$ to 768 unless stated otherwise. This way of designing the Tokenized MLP block helps in encoding meaningful feature information and not contribute much in terms of computation or parameters. 

\section{Experiments and Results}

\noindent \textbf{Datasets:} To make our experiments as close to point-of-care imaging as possible, we pick International Skin Imaging Collaboration (ISIC 2018) \cite{codella2018skin} and Breast UltraSound Images (BUSI) \cite{al2020dataset} datasets to benchmark our results. The ISIC dataset  contains camera-acquired dermatologic images and corresponding segmentation maps of skin lesion regions. The ISIC 2018 dataset consists of 2594 images. We resize all the images to a resolution of $512 \times 512$. BUSI consists of ultrasound images of normal, benign and malignant cases of breast cancer along with the corresponding segmentation maps. We use only benign and mailgnant images which results in a total of 647 images resized to a resolution of $256 \times 256$. 

\noindent \textbf{Implementation Details:} We develop UNeXt using Pytorch framework. We use a combination of binary cross entropy (BCE) and dice loss to train UNeXt. The loss $\mathcal{L}$ between the prediction $\hat{y}$ and the target $y$ is formulated as:
\begin{equation}
    \mathcal{L} = 0.5  BCE(\hat{y}, y) + Dice(\hat{y}, y)
\end{equation}

We use an Adam optimizer with a learning rate of 0.0001 and momentum of 0.9. We also use a cosine annealing learning rate scheduler with a minimum learning rate upto 0.00001. The batch size is set equal to 8. We train UNeXt for a total of 400 epochs. We perform a 80-20 random split thrice across the dataset and report the mean and variance.

\noindent \textbf{Performance Comparison:} We compare the performance of UNeXt with recent and widely used medical image segmentation frameworks. In particular, we compare with convolutional baselines like UNet \cite{ronneberger2015u}, UNet++ \cite{zhou2018unet++} and ResUNet \cite{zhang2018road}. We also compare with very recent transformer baselines like TransUNet \cite{chen2021transunet} and MedT \cite{jose2021medical}. Note that we have focused on comparing against the baselines in terms of segmentation performance (F1 score and IoU) as well as number of parameters, computational complexity (in GFLOPs) and inference time (in ms). 

We tabulate the results in Table \ref{quan}. It can be observed that UNeXt obtains better segmentation performance than all the  baselines with close second being TransUNet. The  improvements are statistically significant with $p < 10^{-5}$. However, the most compelling point to note here is that UNeXt has very less number of computation compared to TransUNet as UNeXt does not have any attention blocks. The computation is calculated in terms of the number of floating point operators (FLOPs). We note that UNeXt has the least GFLOPs of 0.57 compared to TransUNet's 38.52 and UNet's 55.84.  It is also the most light-weight network compared to all baselines. In particular, we note that UNeXt has only 1.58 M parameters compared to 105.32 M parameters of TransUNet. We also present the average inference time while operating on a CPU. Note that we have specifically bench-marked the inference time in CPU instead of GPU as point-of-care devices mostly operate on low-compute power and often do not have the computing advantage of GPU.  We perform feed forward for 10 images of resolution $256 \times 256$ and report the average inference time. The CPU used for bench-marking was an Intel Xeon Gold 6140 CPU operating at 2.30 GHz. It can be noted that we experimented with Swin-UNet \cite{cao2021swin} but found have problems with convergence on small datasets resulting in poor performance. However, Swin-UNet is heavy with 41.35 M parameters and also computationally complex with 11.46 GFLOPs.

\begin{table}[htbp]
\centering
\caption{Performance Comparison with convolutional and transformer baselines.}
\label{quan}

\resizebox{1\columnwidth}{!}{%
\begin{tabular}{c|c|c|c|cc|cc}
\hline
\multirow{2}{*}{Networks} & \multirow{2}{*}{\begin{tabular}[c]{@{}c@{}}Params \\ (in M)\end{tabular}} & \multirow{2}{*}{\begin{tabular}[c]{@{}c@{}}Inference \\ Speed (in ms)\end{tabular}} & \multirow{2}{*}{GFLOPs} & \multicolumn{2}{c|}{ISIC \cite{codella2018skin}}             & \multicolumn{2}{c}{BUSI \cite{al2020dataset}}                                  \\ \cline{5-8} 
                          &                             &                                                                             &                         & \multicolumn{1}{c|}{F1} & IoU         & \multicolumn{1}{c|}{F1} & IoU                             \\ \hline
UNet \cite{ronneberger2015u}                    & 31.13                       & 223                                                                       & 55.84                   & 84.03 $\pm$ 0.87             & 74.55 $\pm$ 0.96  & 76.35 $\pm$ 0.89              & 63.85 $\pm$ 1.12                      \\
UNet++ \cite{zhou2018unet++}                  & 9.16                        & 173                                                                        & 34.65                   & 84.96 $\pm$ 0.71              & 75.12 $\pm$ 0.65  & 77.54 $\pm$ 0.74              & 64.33 $\pm$ 0.75                      \\
ResUNet \cite{zhang2018road}              & 62.74                       & 333                                                                       & 94.56                   & 85.60 $\pm$ 0.68              & 75.62 $\pm$ 1.11  & 78.25 $\pm$ 0.74              & 64.89 $\pm$ 0.83                      \\ \hline
MedT \cite{jose2021medical}                     & 1.60                        & 751                                                                        & 21.24                   & 87.35 $\pm$ 0.18              & 79.54 $\pm$ 0.26  & 76.93 $\pm$ 0.11              & 63.89 $\pm$ 0.55                      \\
TransUNet \cite{chen2021transunet}               & 105.32                      & 246                                                                       & 38.52                   & 88.91 $\pm$ 0.63              & 80.51 $\pm$ 0.72  & 79.30 $\pm$ 0.37              & 66.92 $\pm$ 0.75                      \\ \hline
\textbf{UNeXt}                & \textbf{1.47}                        & \textbf{25}                                                                        & \textbf{0.57}                    & \textbf{89.70 $\pm$ 0.96}           & \textbf{81.70 $\pm$ 1.53} & \textbf{79.37 $\pm$ 0.57}             & \textbf{66.95 $\pm$ 1.22} \\ \hline
\end{tabular}}
\end{table}

\begin{figure}[htbp]
	\centering
	\includegraphics[width=1\linewidth]{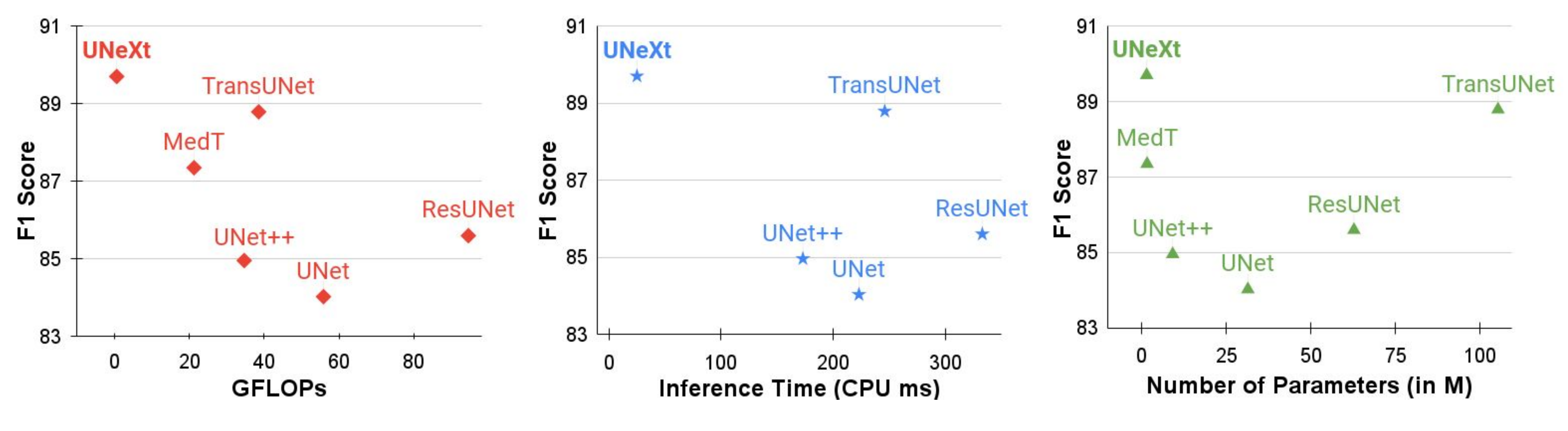}
	
	\vskip-11pt\caption{\textbf{Comparison Charts.} Y-axis corresponds to F1 score (higher the better). X-axis corresponds to GFLOPs, inference time and number of parameters (lower the better). It can be seen that UNeXt is the most efficient network  compared to the others.  }  
	\label{graph}

\end{figure}

\begin{figure*}[h]
  \centering
\begin{tabular}[!t]{ccccccc}
  \includegraphics[trim={0cm 0cm 0cm 0cm},clip,width=0.135\linewidth]{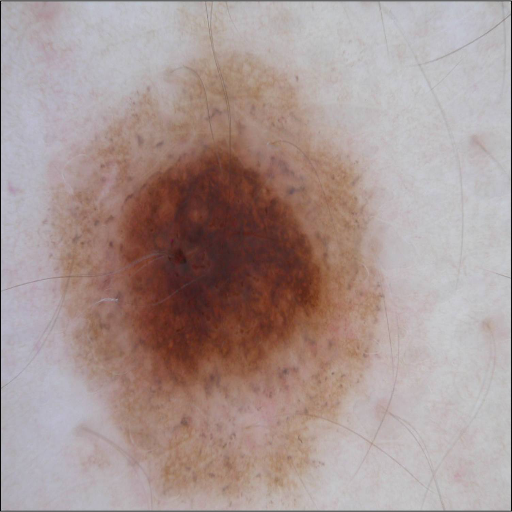} 
  \includegraphics[trim={0cm 0cm 0cm 0cm},clip,width=0.135\linewidth]{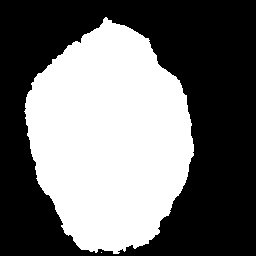}
  \includegraphics[trim={0cm 0cm 0cm 0cm},clip,width=0.135\linewidth]{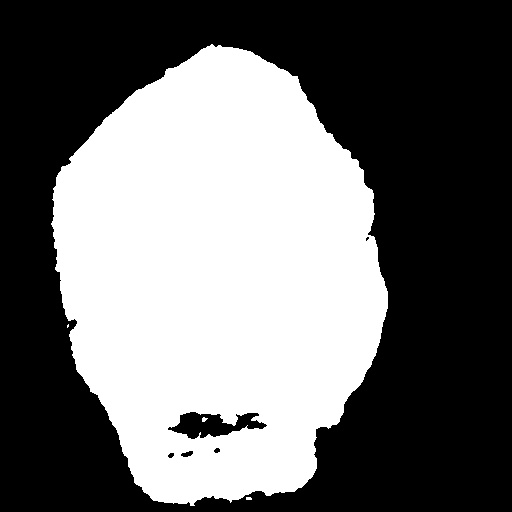}
  \includegraphics[trim={0cm 0cm 0cm 0cm},clip,width=0.135\linewidth]{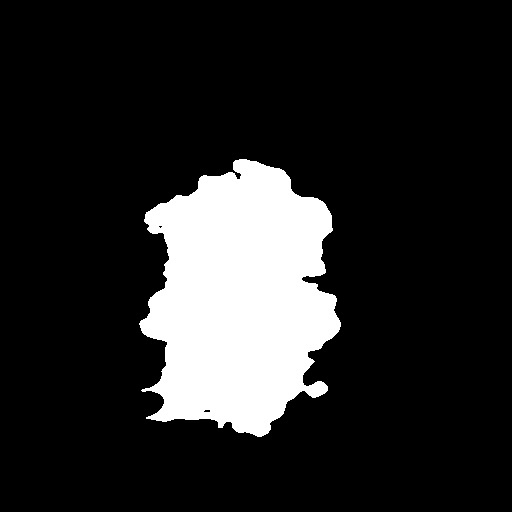}
  \includegraphics[trim={0cm 0cm 0cm 0cm},clip,width=0.135\linewidth]{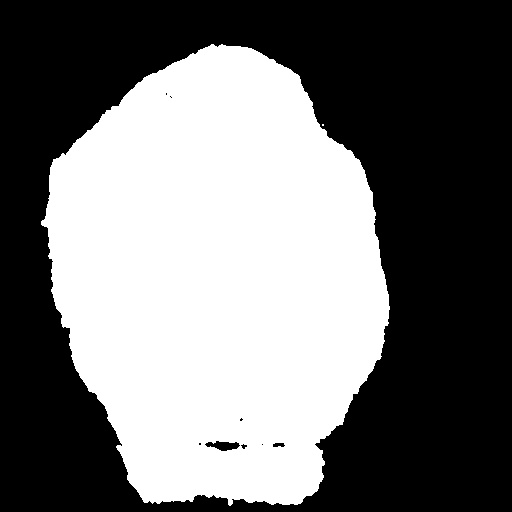}
  \includegraphics[trim={0cm 0cm 0cm 0cm},clip,width=0.135\linewidth]{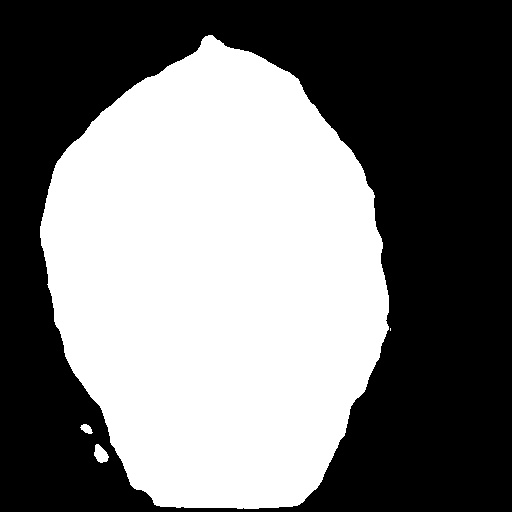}

  \includegraphics[trim={0cm 0cm 0cm 0cm},clip,width=0.135\linewidth]{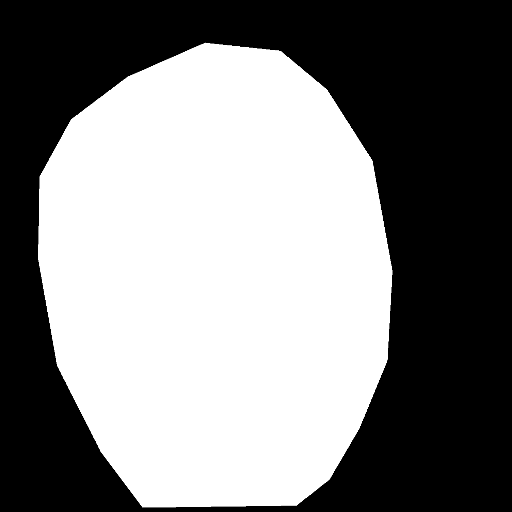} 
  \\
  \includegraphics[trim={0cm 0cm 0cm 0cm},clip,width=0.135\linewidth]{figs/0052_inp.png} 
  \includegraphics[trim={0cm 0cm 0cm 0cm},clip,width=0.135\linewidth]{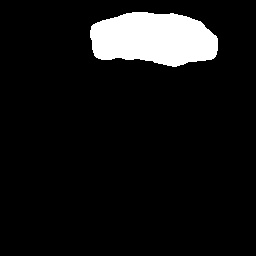}
  \includegraphics[trim={0cm 0cm 0cm 0cm},clip,width=0.135\linewidth]{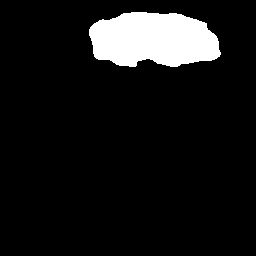}
  \includegraphics[trim={0cm 0cm 0cm 0cm},clip,width=0.135\linewidth]{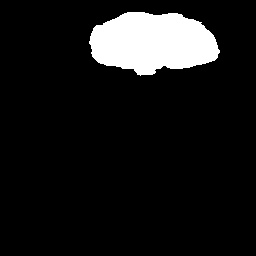}
  \includegraphics[trim={0cm 0cm 0cm 0cm},clip,width=0.135\linewidth]{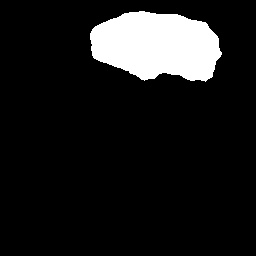}
  
  \includegraphics[trim={0cm 0cm 0cm 0cm},clip,width=0.135\linewidth]{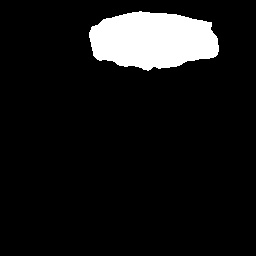} 
 
  \includegraphics[trim={0cm 0cm 0cm 0cm},clip,width=0.135\linewidth]{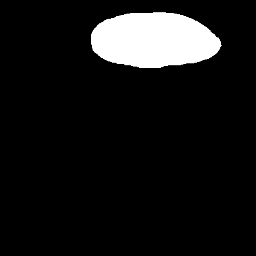} 
  \\
  
(a)\hskip38pt (b)\hskip38pt (c)\hskip38pt (d)\hskip38pt (e)\hskip38pt (f)\hskip38pt (g)

\end{tabular}

\caption{\textbf{Qualitative comparisons.} Row 1 - ISIC dataset, Row 2 - BUSI dataset. (a) Input. Predictions of (b) UNet (c) UNet++ (d) MedT (e) TransUNet (f) UNeXt and (g) Ground Truth. }

\label{qual}
\end{figure*}

In Figure \ref{graph}, we plot the comparison  charts of F1 score vs. GLOPs, F1 score vs. Inference time and F1 Score vs. Number of Parameters. The F1 score used here corresponds to the ISIC dataset. It can be clearly seen from the charts that UNeXt and TransUNet are the best performing methods in terms of the segmentation performance. However, UNeXt clearly outperforms all the other networks in terms of computational complexity, inference time and number of parameters which are all important characteristics to consider for point-of-care imaging applications. In Figure \ref{qual}, we present sample qualitative results of UNeXt along with other baselines. It can be observed that UNeXt produces competitive segmentation predictions compared to the other methods.

\section{Discussion}

\noindent \textbf{Ablation Study:} We conduct an ablation study (shown in Table \ref{abl}) to understand the individual contribution of each module in UNeXt. We first start with the original UNet and then just reduce the number of filters to reduce the number of parameters and complexity. We see a reduction of performance with not much reduction in parameters. Next, we reduce the depth and use only a 3-level deep architecture which is basically the Conv stage of UNeXt. This reduces the number of parameters and complexity significantly but also reduces the performance by ~4\%. Now, we introduce the tokenized MLP block which improves the performance significantly while increasing the complexity and parameters by a minimal value. Next, we add the positional embedding method using DWConv as in \cite{xie2021segformer} and see some more improvement. Next, we add the shifting operation in the MLPs and show that shifting the features before tokenizing improves the performance without any addition to parameters or complexity. As the shift operation does not contribute to any addition or multiplication it does not add on to any FLOPs. We note that shifting the features across both axes results in the best performance which is the exact configuration of UNeXt with  minimal parameters and complexity. Note that all of the above experiments were conducted using a single fold of the ISIC dataset.

\begin{table}[]
\centering
\caption{Ablation Study.}

\resizebox{1\columnwidth}{!}{%
\begin{tabular}{c|c|c|c|c|c}
\hline
Network                                 & Params & Inf. Time (in ms) & GFLOPs & F1 & IoU \\ \hline
Original UNet                               &  31.19      & 223    & 55.84        & 84.03    & 74.55    \\
Reduced UNet                               & 7.65        & 38    & 9.36       & 83.65   & 72.54    \\
Conv Stage                  & 0.88       & 9    &  0.36       & 80.12    & 67.75     \\
Conv Stage  + Tok-MLP w/o PE                    & 1.46       & 22    & 0.57        & 88.78    & 79.32    \\
Conv Stage + Tok-MLP + PE               &  1.47      & 23    & 0.57       & 89.25   & 80.76    \\
Conv Stage + Shifted Tok-MLP (W) + PE   & 1.47       & 24    & 0.57        & 89.38    & 82.01     \\
Conv Stage + Shifted Tok-MLP (H) + PE   & 1.47      & 24    & 0.57       & 89.25     & 81.94     \\
Conv Stage + Shifted Tok-MLP (H+W) + PE & 1.47        & 25    & 0.57        & 90.41    & 82.78     \\ \hline
\end{tabular}
}
\label{abl}
\end{table}

\noindent \textbf{Analysis on number of channels:} The number of channels is a main hyperparamter of UNeXt which affects the number of parameters, complexity and the performance of the network. In Table \ref{analysis}, we conduct experiments on single fold of ISIC to show two more different configurations of UNeXt. It can be observed that increasing the channels (UNeXt-L) further improves the performance while adding on to computational overhead. Although decreasing it (UNeXt-S) reduces the performance (the reduction is not drastic) but we get a very lightweight model.

\begin{table}[]
\centering
\caption{Analysis on the number of channels.}
\label{analysis}

\resizebox{0.9\columnwidth}{!}{%
\begin{tabular}{c|c|c|c|c|c|c|c|c|c|c}
\hline 
Network & C1 & C2 & C3  & C4  & C5  & Params & Inf. Speed (in ms) & GFLOPs & F1 & IoU \\ \hline
UNeXt-S & 8  & 16 & 32  & 64  & 128 & 0.32       & 22            &  0.10      & 89.62   & 81.40     \\
UNeXt-L & 32 & 64 & 128 & 256 & 512 & 3.99       & 82           & 1.42        & 90.65    & 83.10     \\
UNeXt & 16 & 32 & 128 & 160 & 256 & 1.47       & 25           & 0.57       & 90.41    & 82.78     \\ \hline
\end{tabular}
}
\end{table}



\noindent \textbf{Difference from MLP-Mixer:} MLP-Mixer uses an all-MLP architecture for image recognition. UNeXt is a convolutional and MLP-based network for image segmentation. MLP-Mixer focuses on channel mixing and token mixing to learn a good representation. In contrast, we extract convolutional features and then tokenize the channels and use a novel tokenized MLPs using shifted MLPs to model the representation. It is worthy to note that we experimented with MLP-Mixer as encoder and a normal convolutional decoder. The performance was not optimal for segmentation and it was still heavy with  around 11 M parameters.

\section{Conclusion}

In this work, we have proposed a new deep network architecture UNeXt for medical image segmentation focussed for point-of-care applications. UNeXt is a convolutional and MLP-based architecture where there is an initial conv stage followed by MLPs in the latent space. Specifically, we propose a tokenized MLP block with shifted MLPs to efficiently model the representation with minimal complexity and parameters. We validated UNeXt on multiple datasets where we achieve faster inference, reduced complexity and less number of parameters while also achieving the state-of-the-art performance.

\bibliographystyle{splncs04}
\bibliography{fast}
	
\end{document}